\documentclass[12pt]{iopart}
\usepackage{epsf}

\def\GeV{\hbox{$\;\hbox{\rm GeV}$}}
\def\MeV{\hbox{$\;\hbox{\rm MeV}$}}
\newcommand{\Rp}{\mbox{$\not \hspace{-0.15cm} R_p$}}
\newcommand{\gsim}{\raisebox{-0.5mm}{$\stackrel{>}{\scriptstyle{\sim}}$}}

\begin{document}
% Journal identifier can be put here if required, e.g.
%\jl{14}

\begin{flushright}
WUE-ITP-98-056 \\
DAPNIA/SPP 98-23
\end{flushright}

\title{High $Q^2$ Physics at HERA and 
       Searches for New Particles}

\author{T.~Matsushita$^1$, E.~Perez$^2$, R.~R\"uckl$^3$}
% \address{$^1$ KEK/IPNS, Tokyo, Japan, ZEUS Collaboration}
\address{$^1$ Department of Physics, University of Oxford, U.K., ZEUS
              Collaboration}
\address{$^2$ CEA-Saclay, DSM/DAPNIA/Spp, France, H1 Collaboration}
\address{$^3$ Institut f\"ur Theoretische Physik,
              Universit\"at W\"urzburg, Germany}

\begin{abstract}
Preliminary results from H1 and ZEUS 
on Deep Inelastic Scattering (DIS)
at high momentum transfer squared $Q^2$ are presented.
% ===> TM
% Used are all available $e^+ p$ data accumulated
% between 1994 and 1997, corresponding to integrated
% luminosities of $\simeq 37 {\mbox{pb}^{-1}}$ (H1) and
% $\simeq 46 {\mbox{pb}^{-1}}$ (ZEUS).
%
% H1 and ZEUS experiments collected $e^+p$ data
% corresponding to integrated luminosities of
% $37 {\mbox{pb}^{-1}}$ and $47 {\mbox{pb}^{-1}}$
% respectively, during the years 1994 to 1997.
Used are all available $e^+ p$ data accumulated
by the H1 and ZEUS experiments between 1994 and 1997,
corresponding to integrated luminosities of
$37 {\mbox{pb}^{-1}}$ and $47 {\mbox{pb}^{-1}}$,
respectively.
% ===> End TM
The anomalies observed at high $Q^2$ in the 1994 to 1996
data still remain, though with less significance.
Since this high $Q^2$ domain represents a new frontier in DIS,
the same data are used to search for new particles possessing
direct couplings to lepton-quark pairs.
Assuming that the slight excess of events observed in Neutral Current DIS
is due to a statistical fluctuation, preliminary limits
on the production of leptoquarks and of squarks in R-parity violating 
MSSM are presented.
\end{abstract}

\vspace*{10cm}
\begin{flushleft}
 {\small{
       Contribution to the 3rd UK Phenomenology
       Workshop on HERA Physics, St. John's College,
       Durham, UK, September 1998.}}
\end{flushleft}

\maketitle

%============================
\section{Introduction}
%============================

Deep Inelastic Scattering (DIS) of leptons by nucleons at 
% ===> TM Dec 16
% the $ep$ collider HERA offers the unique possibility to probe
the $ep$ collider HERA offers an unique possibility to probe
% ===> End
the proton at very small distances ($\simeq 10^{-16} {\mbox{cm}}$).
The domain of very high momentum transfer squared $Q^2$ is 
also a window where first signs
of physics beyond the Standard Model (SM) 
may manifest themselves. 

Using 1994 to 1996 data, corresponding to
a total integrated luminosity of about $35 {\mbox{pb}}^{-1}$,
H1~\cite{H1HIQ2} and ZEUS~\cite{ZEUSHIQ2} experiments
% ===> TM Dec 16
% reported the observation of
reported an observation of
% ===> End
an excess of events in DIS at $Q^2 > 15000 \GeV^2$. This
finding stimulated a lot of experimental and theoretical activity.
Data accumulated during 1997 allowed to increase the statistics
by roughly a factor $2.5$. 
The measurements of DIS at high $Q^2$ based on all available
$e^+ p$ data are presented and compared to the SM
expectations in section~\ref{sec:dis}.
The same data are then used to search for new
particles possessing direct couplings to lepton-quark pairs, 
namely leptoquarks and squarks 
in SUSY models where R-parity is violated (\Rp).
Preliminary results on leptoquark production are shown in
section~\ref{sec:leptoquarks}, together with some remarks
on particular phenomenological aspects at HERA.
Analogous constraints on resonant \Rp-squarks
production are presented in section~\ref{sec:rpvsusy}.

%==========================================================
\section{Deep Inelastic Scattering at high $Q^2$ at HERA}
%==========================================================

\label{sec:dis}

%----------------------------------------------------------
\subsection{Comparison of DIS data with the SM expectations}
%----------------------------------------------------------

\begin{figure}[htb]
 \begin{center}
 \hspace*{-2cm}\begin{tabular}{p{0.40\textwidth}p{0.60\textwidth}}
 \vspace*{-8cm}\caption
         { \label{fig:h1figq2}
              (a) $Q^2$ distribution of the observed NC DIS 
              events in comparison to the SM expectation
              (histogram);
              (b) ratio of the observed and expected number of
              events as a function of $Q^2$;
              (c,d) same as (a,b) but for 
              the 1997 data alone. } &
 \epsfxsize=0.7\textwidth
 \epsfbox{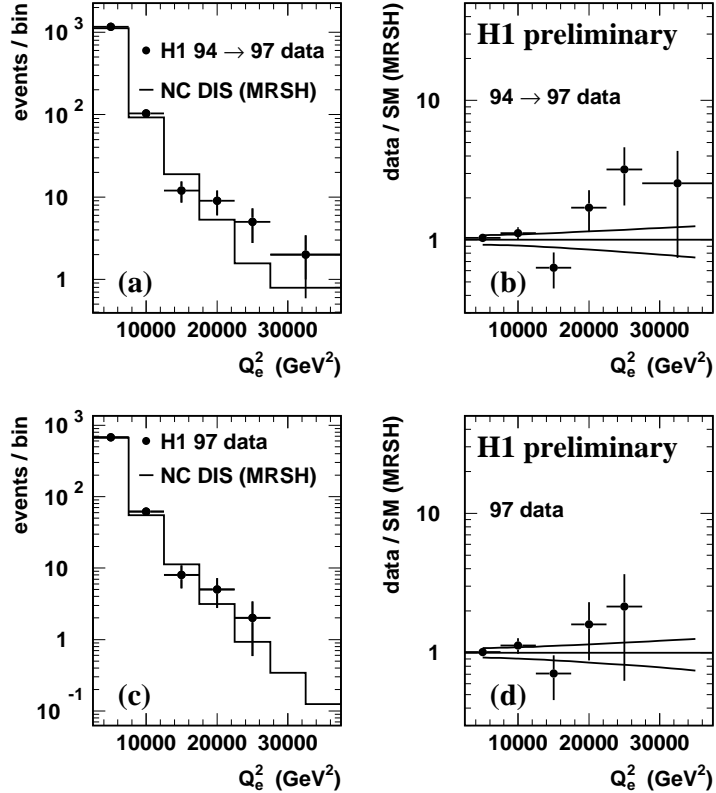}
 \end{tabular}
 \end{center}
\end{figure}

The full available $e^+ p$ data have been used by H1 and ZEUS
experiments to update the measurements of Neutral Current (NC)
% ===> TM Dec 16
% and Charged Current (CC) DIS presented in~\cite{H1HIQ2,ZEUSHIQ2}.
and Charged Current (CC) DIS presented in~\cite{H1ELAN,ZEUSNC}.
% ===> End
%
% ===> TM
% The measured $Q^2$ distribution for the H1 NC selection~\cite{H1HIQ2}
% ===> TM Dec 16
% The distribution in $Q^2$ of the NC DIS events measured by H1~\cite{H1HIQ2}
The distribution in $Q^2$ of the NC DIS events measured by H1~\cite{H1ELAN}
% ===> End
is shown in Fig.~\ref{fig:h1figq2}a (all data,
${\cal{L}} = 37.4 {\mbox{pb}}^{-1}$) and Fig.~\ref{fig:h1figq2}c
(1997 data alone, ${\cal{L}} = 22.8 {\mbox{pb}}^{-1}$),
in comparison to the SM expectation.
Also shown in Fig.~\ref{fig:h1figq2}b and Fig.~\ref{fig:h1figq2}d are
the ratios of these measurements to theory.
The curves above and below unity specify the $\pm 1\sigma$ levels
determined from the combination
of statistical and systematic errors of the NC DIS expectation. 
These errors are
dominated by the uncertainty on the electromagnetic energy scale of the
calorimeter and vary between $8.5 \%$ at low $Q^2$ and about $30 \%$ 
at the highest $Q^2$.
%
%------------------------------------------------------
%   FIGURE 1 : H1 Q2 spectrum
%------------------------------------------------------
%
% ===> TM :
% It is seen that the NC DIS expectation agrees well with the data at 
The NC DIS expectation agrees well with the data at
% ===> End TM
$Q^2 < 10000 \GeV^2$, while at larger $Q^2$ deviations are
observed with a slight deficit of events between 10000 $\GeV^2$
and $15000 \GeV^2$ and a slight excess at $Q^2$ $\gsim$ $15000 \GeV^2$.
The 1997 data alone suggest similar deviations, although with
a more marginal significance.
%
% ===> TM :
% Table~\ref{tab:tabq2} summarizes the numbers of events 
% with $Q^2$ above various thresholds $Q^2_{min}$ observed by H1 and ZEUS
% and expected in SM.
Table~\ref{tab:tabq2} summarizes the numbers of events
observed by both experiments and expected in the SM with
$Q^2$\footnote{
              $Q^2_e$ and $Q^2_{2\alpha}$ mean $Q^2$
              reconstructed by the electron and double angle method,
              respectively.} 
above various thresholds $Q^2_{min}$.
% ===> End TM
%
%------------------------------------------------------
%  TABLE 1 : Events rates (H1,ZEUS), NC DIS
%
\begin{table*}[htb]
 \begin{center}
 \hspace*{-2cm}\begin{tabular}{p{0.40\textwidth}p{0.60\textwidth}}
         \vspace*{-2cm}\caption {\label{tab:tabq2}
            Number of observed ($N_{obs}$) and expected 
           ($N_{exp}$) NC DIS events with momentum transfer squared
           $Q^2$ above given thresholds $Q^2_{min}$. } &
\begin{tabular}{||c|c|c|c||}
 \hline \hline
 $Q^2_{min}$                         & Exp.        & $N_{obs}$       & $N_{exp}$  \\ \hline
 $ Q^2_e > 15000 \GeV^2$             &  H1         & 22              & $14.8 \pm 2.1 $ \\ 
 $ Q^2_{2 \alpha} > 15000 \GeV^2$    & ZEUS        & 20              & $ 17 \pm 2$ \\ \hline
 $ Q^2_e > 25000 \GeV^2$             &  H1         & 6               & $1.6 \pm 0.3$ \\ \hline
 $ Q^2_{2 \alpha} > 35000 \GeV^2$    & ZEUS        & 2               & $0.29 \pm 0.02$ \\
 \hline \hline
\end{tabular}
\end{tabular}
 \end{center}
\end{table*}
%------------------------------------------------------
%
% ===> TM 
At $Q^2 > 35000 \GeV^2$, ZEUS observes two events while
0.29 are expected. These events were reported previously~\cite{ZEUSHIQ2}.
In the 1997 data,
no additional events were found in this $Q^2$ domain. 
% At $Q^2 > 35000 \GeV^2$, ZEUS observes two events while
% 0.29 are expected. However no additional events were found in this $Q^2$
% domain in the 1997 data.

%---------------------------------------------------
\subsection{Differential Cross Sections}
%---------------------------------------------------

% ===> TM
% The observed $Q^2$ distributions can be converted into differential
% cross sections~\cite{H1ELAN,ZEUSNC,ZEUSCC}.
Differential cross sections in terms of $Q^2$ are
derived by both experiments~\cite{H1ELAN,ZEUSNC,ZEUSCC}.
% ===> End TM
For that purpose, the number of events in each $Q^2$
bin is corrected for acceptance and migration, using Monte Carlo
simulation, and for photon radiation and electroweak effects, 
using the HERACLES~\cite{HERACLES} program. 
%
%------------------------------------------------------
%   FIGURE 2 : H1 + ZEUS NC/CC Cross sections
\begin{figure}[htb]
 \begin{center}
  \begin{tabular}{p{0.40\textwidth}p{0.60\textwidth}}
         \vspace*{-6cm}\caption
         { \label{fig:dsdq2}
          Differential cross sections $d \sigma / d Q^2$ for NC and CC
               DIS, measured by H1 (squares) and ZEUS (points), using
               all available $e^+ p$ data.
          The curves represent the SM predictions. } &
 \epsfxsize=0.5\textwidth
 \epsfbox{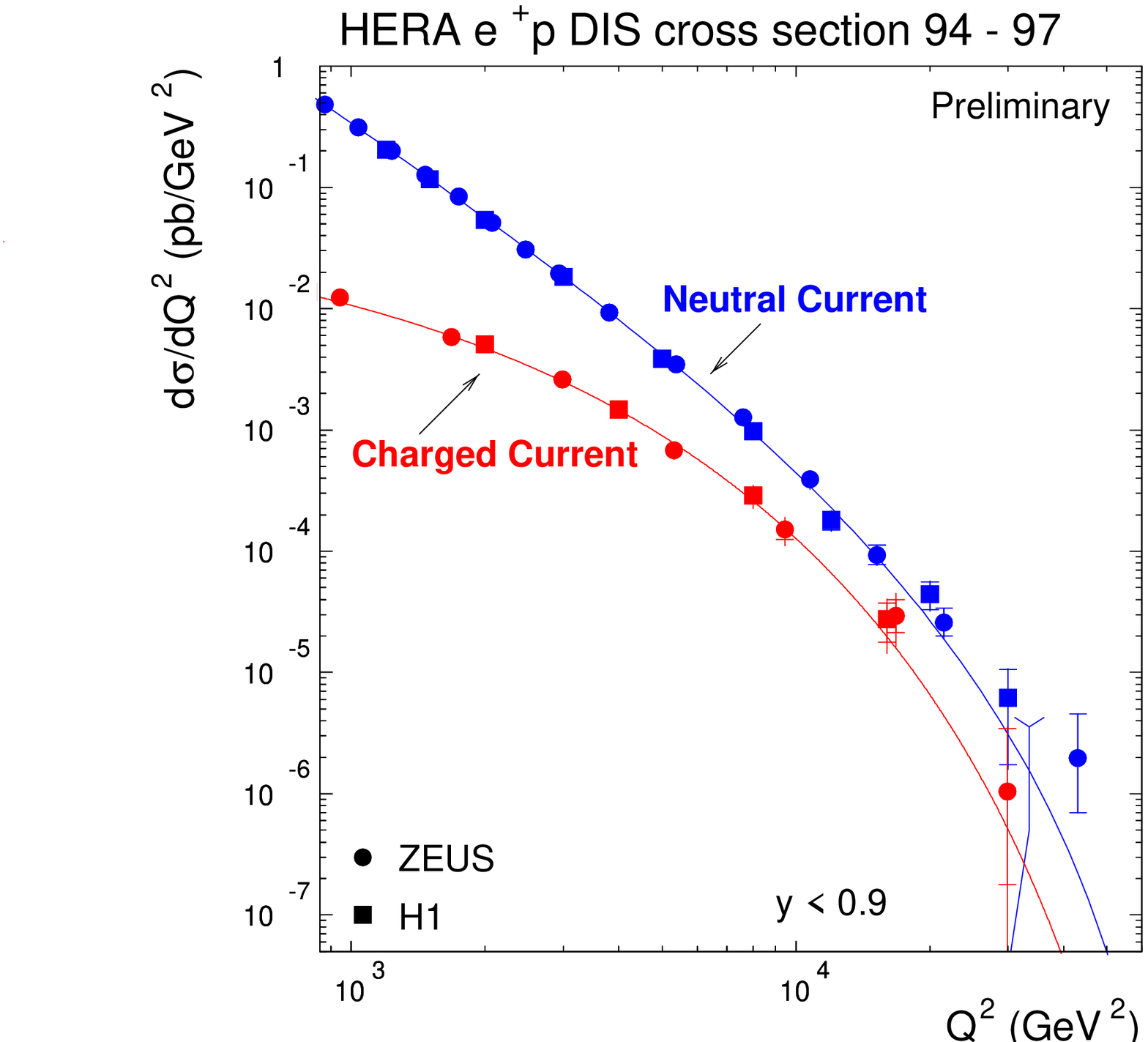}
  \end{tabular}
  \end{center}
 \end{figure}
%------------------------------------------------------
%
Fig.~\ref{fig:dsdq2} shows the NC and CC cross sections 
$d \sigma / d Q^2$ 
% resulting from the H1 and ZEUS data.
obtained by H1 and ZEUS.

\begin{figure}
 \begin{center}
  \begin{tabular}{cc}
   \epsfxsize=0.5\textwidth
   \epsfbox{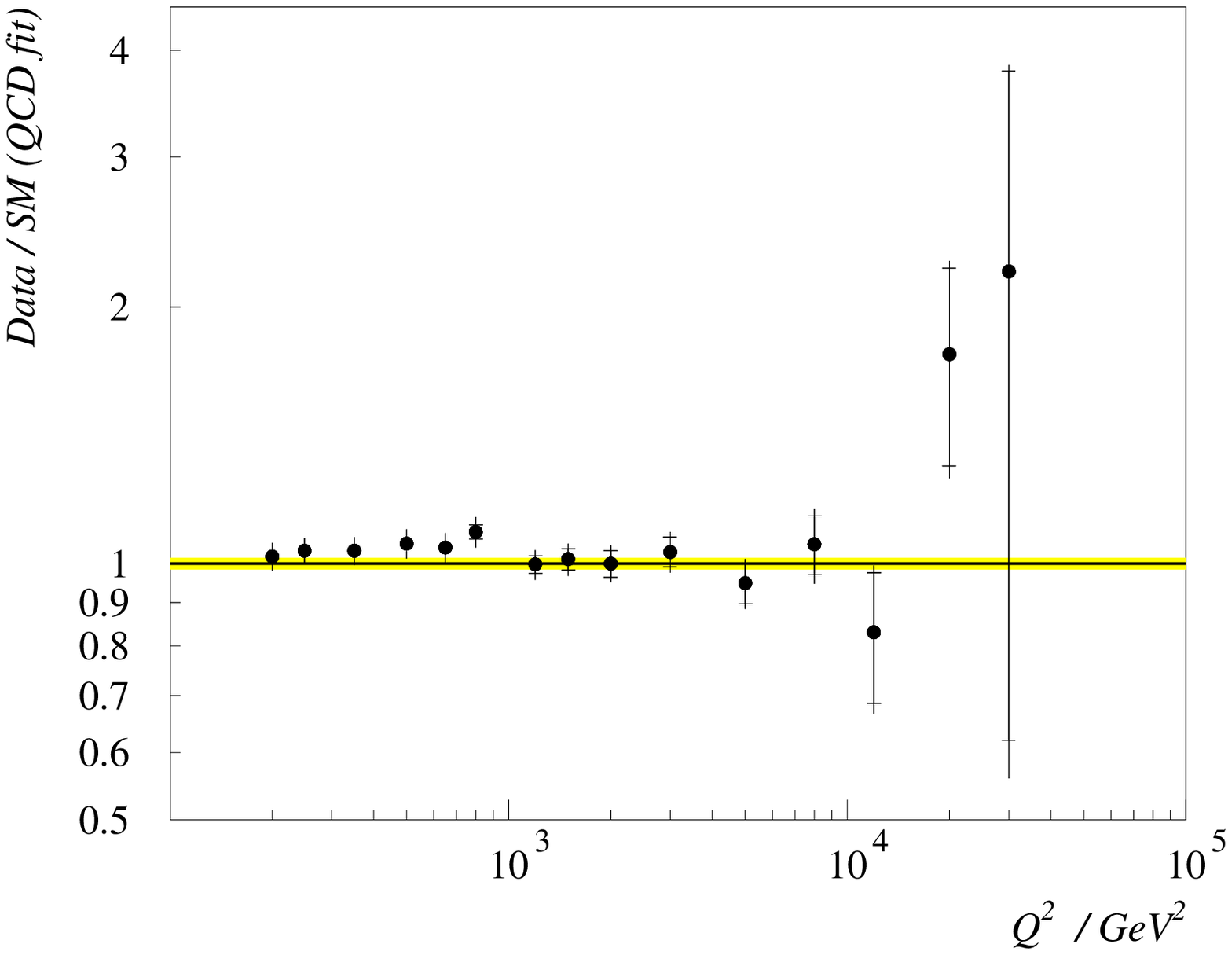} &
   \epsfxsize=0.5\textwidth
   \epsfbox{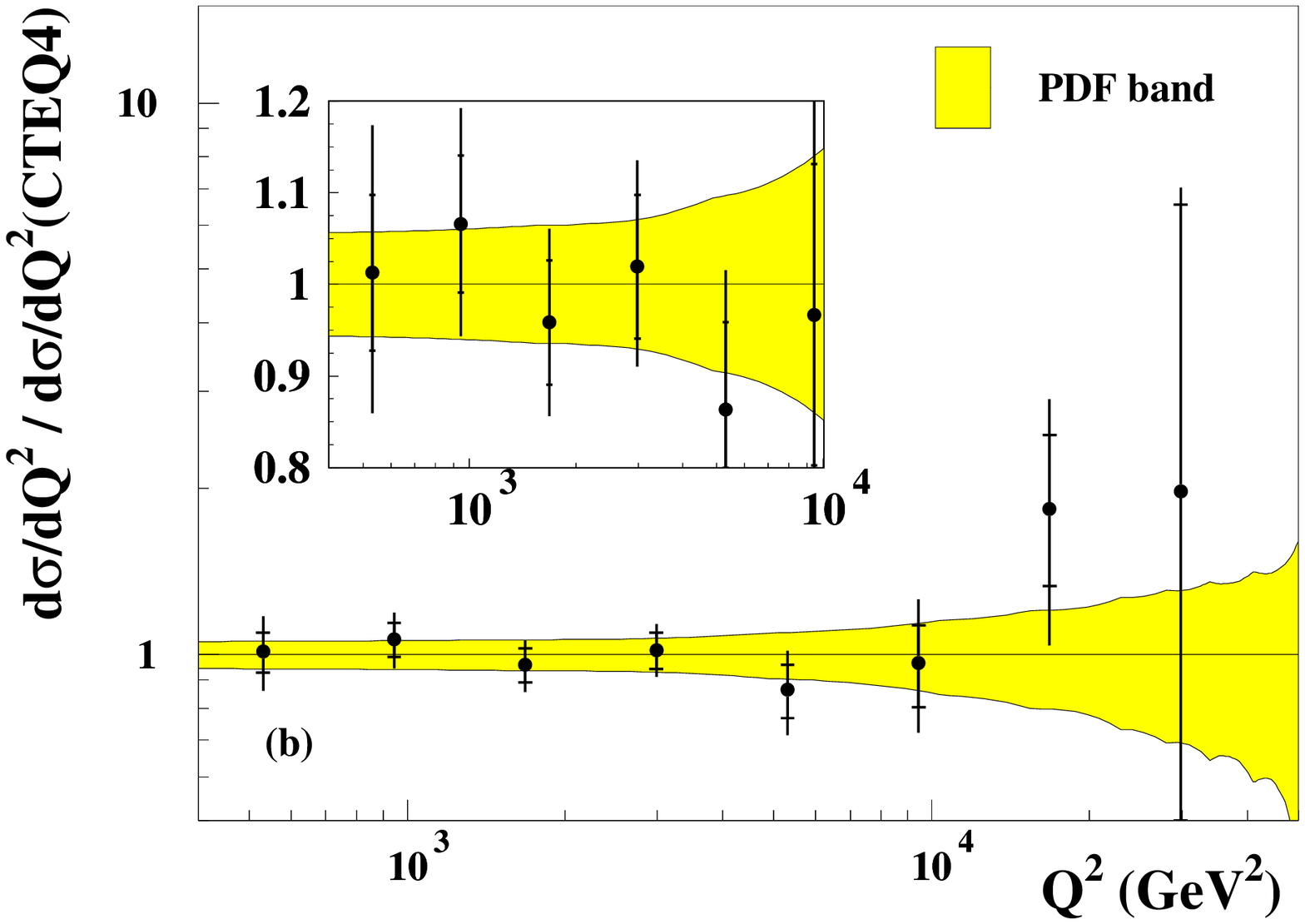}
   \end{tabular}
  \caption[]{ {\label{fig:dsdq2ratio} Left: ratio of the NC DIS cross section 
                measured by H1 to the SM prediction.
                Right: ratio of the CC DIS cross section measured by ZEUS
                to the SM prediction. On both figures, the inner
                bars represent the statistical error and the outer ones
                the total error.}}
                
  \end{center}
 \end{figure}

For NC DIS, the rapid fall off of $d\sigma /dQ^2$ by about
6 orders of magnitude 
% on the whole agrees with the SM prediction.
is on the whole well described by the SM prediction.
However, as already mentioned, the measurements slightly undershoot
the SM expectation at $Q^2 \simeq 10000 \GeV^2$, while a slight
excess of events is observed in the two highest $Q^2$ bins.
The ratio of the experimental (H1) and theoretical NC cross section
is shown in Fig.~\ref{fig:dsdq2ratio} (left).
Here, the SM prediction is obtained from a NLO QCD fit
performed by H1~\cite{H1ELAN}, which includes the new H1
high $Q^2$ data as well as H1 and fixed target charged lepton-nucleon
data at lower $Q^2$.
% The ratios of H1 NC DIS measured cross section to the SM prediction
% is shown in Fig.~\ref{fig:dsdq2ratio} (left). Here, SM expectation refers
% to a NLO QCD fit performed by H1~\cite{H1ELAN}, 
% including the new H1 high $Q^2$ data as well
% as H1 and fixed target charged lepton-nucleon data at lower $Q^2$.
The inner error bars on the data points correspond to the statistical errors,
while the outer ones indicate the full error, the typical 
total systematic error being about $4 \%$.

For CC DIS, the Standard Model also provides a very good description
of the data over most of the kinematic range as can be seen
in Fig.~\ref{fig:dsdq2}.
%
%------------------------------------------------------
%   FIGURE 3  : H1 Ratio ds/dq2 NC + ZEUS Ratio ds/dq2 CC
%------------------------------------------------------
%
%
% ===> TM :
% For CC DIS, the Standard Model provides a very good description
% of the data over most of the kinematic range. 
% Only at the highest $Q^2$ the data have a tendency
% to lie above SM prediction, as can be seen in
% Fig.~\ref{fig:dsdq2ratio} (right) showing the ratio of 
% the measured (ZEUS) to the theoretical (CTEQ4) CC cross section.
% In this figure,
% the shaded region visualizes the uncertainty in the SM prediction due to
% uncertainties in the partons densities.
% As before, the outer error bars on the data points 
% correspond to the total error,
% the main source of systematic error being the uncertainty on the calorimeter
% energy scale.
% The latter leads to an uncertainty on the measured cross section 
% of $5 -10 \%$ in most of the $Q^2$ range, except at the highest
% $Q^2$, where the error rises to about $50 \%$~\cite{ZEUSCC}.
Only at the highest $Q^2$ the data
show a tendency to lie above the SM prediction.
The ratio of the measured (ZEUS) to the theoretical (CTEQ4) 
$d\sigma /dQ^2$ distribution for CC DIS is shown in 
Fig.~\ref{fig:dsdq2ratio} (right).
The shaded region indicates the theoretical uncertainty due
to the uncertainties in the partons densities.  
As before, the outer error bars on the data points
correspond to the total error. The main source of systematic error
is the calorimeter energy scale.
It leads to an error on the measured cross section
of $5 -10 \%$ in most of the $Q^2$ range, except at the highest
$Q^2$, where the error rises to about $50 \%$~\cite{ZEUSCC}.
% The SM provides a very good description of
% the data except for the highest $Q^2$ bins where the data show a
% tendency to lie above the prediction.
% ===> End TM

% ===> EP : Refer to dsigma/dx with A.Bodek's d/u
% ZEUS CC differential cross section $d \sigma / d x$ has also
% been measured and compared to the SM prediction where
% CTEQ4 parametrization has been modified using a
% larger $d/u$ ratio at high $x$~\cite{BODEK}. 
The CC differential cross section $d \sigma/dx$ has also been
measured by ZEUS and compared to the SM prediction.
% ===> TM Dec 16
% The tendency to observe the data above the prediction can be seen.
One can see some tendency in the data to lie above the expectation.
% ===> End
If the CTEQ4
parametrization of parton densities is modified to give a larger
$d/u$ ratio at high $x$~\cite{BODEK},
the excess observed at high $x$ relative to CTEQ4 becomes less
significant~\cite{WG1}.
% ===> End EP 

%==========================================================
\section{Leptoquark Searches}
%==========================================================

\label{sec:leptoquarks}

%----------------------------
\subsection{Phenomenology}
%----------------------------

Leptoquarks (LQ) are scalar or vector bosons which carry 
both lepton and quark quantum numbers. They appear in many
extensions of the SM. At HERA, LQs may be produced resonantly,
via the fusion of the incident lepton with a quark or
antiquark coming from the proton.

%........................................
%\subsubsection{The considered model}
%........................................
%
% {\bf{The considered model: }}
A reasonable phenomenological framework is provided by 
the BRW model~\cite{BRW}. This model is based on
the most general Lagrangian that is invariant under
$SU(3) \times SU(2) \times U(1)$, respects lepton and baryon number
conservation, and incorporates dimensionless couplings of
LQs to left- and/or right-handed fermions.
Under these assumptions LQs can be classified into
10 isospin multiplets, half of which carry fermion number $F=0$
and couple to $e^+ + q$, while the other half carries
$F=2$ and couples to $e^+ + \bar{q}$.
The $F=0$ ($F=2$) class contains 2 (3) scalar and 3 (2) vector
multiplets. In order to avoid stringent bounds
from low energy experiments which would put
LQs outside the reach of present day colliders, couplings
are allowed to only one combination of fermion chiralities,
and only within a single family.
%
% The BRW model~\cite{BRW} considers the most general Lagrangian invariant under
% $SU(3) \times SU(2) \times U(1)$, respecting lepton and baryon numbers
% conservation, with flavor diagonal dimensionless couplings.
% LQ have been classified into 10 isospin multiplets, with couplings to
% left- or right-handed fermions, among which there are 5 scalar leptoquark
% isospin families. They carry fermion number $F=0$ (if coupling to
% $e^+ + q$) or $F=2$ (if coupling to $e^+ + \bar{q}$).
%
The couplings are generically called $\lambda$
and are unknown parameters of the model. 
In the above model, first generation LQs only
decay into $e + q$ and $\nu_e + q$ with fixed branching ratios,
$\beta (eq) + \beta (\nu q) = 1$.
We will use the nomenclature of~\cite{HERAWS} to label 
the different LQ states.

\begin{figure}[htb]
 \begin{center}
  \hspace*{-2cm}\begin{tabular}{p{0.40\textwidth}p{0.60\textwidth}}
         \vspace*{-3cm}\caption
         { \label{fig:lqprod}
            Illustrative diagrams for resonant LQ production 
            in the $s$--channel (a) and $u$--channel
            LQ exchange (b). } &
 \epsfxsize=0.65\textwidth
 \epsfbox{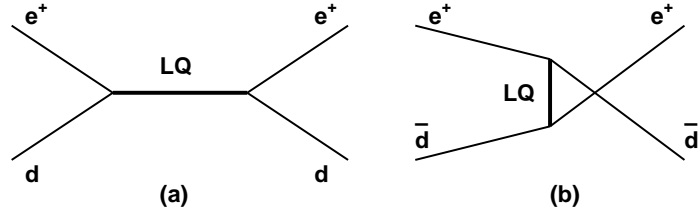}
 \end{tabular}
 \end{center}
 \vspace*{-1.5cm}
\end{figure}

In the following analysis, it is assumed that only one of the LQ multiplets is
produced at a time, and that the states within a given isospin
multiplet are degenerate in mass.
At HERA, LQs can be resonantly produced in the $s$--channel and
exchanged in the $u$--channel. This is illustrated by the
diagrams in Fig.~\ref{fig:lqprod}.
%
%------------------------------------------------------
%   FIGURE  4 : LQ s- and u-channels
%------------------------------------------------------
%
Obviously, the $s$-- and $u$--channel processes do not 
% ===> TM Dec 16
% interfere with each other, but each of these channels interfers with 
interfere with each other, but each of these channels interferes with 
% ===> End
the SM $\gamma$ and $Z$ exchange. The relevant
matrix elements can be found in~\cite{BRW}\footnote{
   The signs of the interference terms between standard
   model gauge boson contributions and fermion number $F=0$
   leptoquark contributions given in~\cite{BRW} are
   incorrect. The correct signs can be inferred from~\cite{KALINOW}
   and amount to multiplying the right-hand-side of
   Eqs (14d,e) and (17d,e) by (-1). }.
%  
% The decay width of a LQ into a lepton and a quark, 
% $\Gamma^S_{LQ} = \lambda^2 M / 16 \pi$ 
% and $\Gamma^V_{LQ} = \lambda^2 M / 16 \pi$ is generally small.
%

For the allowed strength of the couplings and in the mass range
interesting for HERA, the decay width of a LQ into a lepton and a quark
% ===> EP :
is generally small~: 
$ \Gamma_S = 3/2 \Gamma_V = \lambda^2 M / 16 \pi \simeq 40 \MeV$ 
for $M = 200 \GeV$ and $\lambda = 0.1$, $S$ and $V$ referring
to scalar and vector states, respectively.
% ===> End EP
In the narrow-width approximation (NWA), the resonant production cross
section is given by 
$ \sigma_{NWA} = \frac{\pi}{4s} \lambda^2 q(x)$, $q(x)$ being
the probability to find the relevant quark $q$ in the proton with
a fraction $x = M^2 / s$ of the proton momentum. 
In the above, $\sqrt{s} \simeq 300 \GeV$ is the collision
energy in the $ep$ centre of mass and $M$ is the LQ mass.
%
%..........................................................
%\subsubsection{Effect of the non-vanishing width of the LQ}
%..........................................................
%
%{\bf{Effect of the non-vanishing width of the LQ :}}

For high mass LQs, i.e. at high $x$, the parton densities in the
proton are falling very steeply.
At the same time, the LQ width increases with $M$ leading to
important tails at smaller values of $x$. As a result, the convolution of 
the parton density with the corresponding Breit-Wigner distribution gives a 
considerable larger cross section than the NWA.
%
%------------------------------------------------------
%   FIGURE 5  : Effect of narrow width approximation
\begin{figure}
 \begin{center}
  \hspace*{-2cm}\begin{tabular}{p{0.40\textwidth}p{0.60\textwidth}}
         \vspace*{-4cm}\caption
         { \label{fig:appronwa}
              Ratio of the cross section $\sigma_{\Gamma}$ 
              for $s$--channel LQ
              production and finite LQ width $\Gamma$ to the narrow-width
              approximation. } &
 \epsfxsize=0.7\textwidth
 \epsfbox{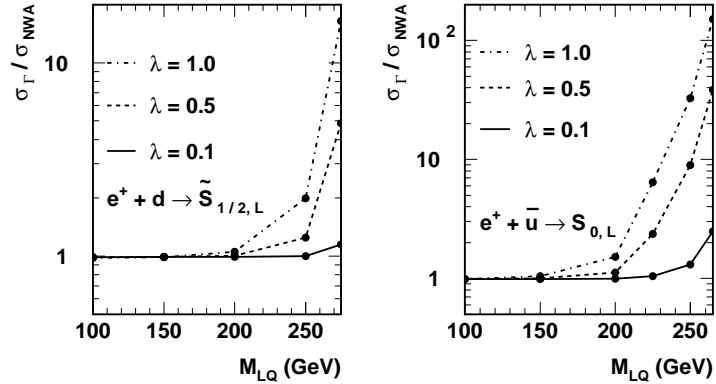}
 \end{tabular}
 \end{center}
 \vspace*{-1.5cm}
\end{figure}
%------------------------------------------------------
%
This is quantified in Fig.~\ref{fig:appronwa}, for the typical
range of couplings which can be probed at HERA with the current
luminosity.
One can see that
for $F=0$ and $\lambda = 0.5$ the deviation from NWA becomes
noticeable at $M > 250 \GeV$, while for $F=2$ the 
finite width becomes non-negligible already at $M > 200 \GeV$.
%
% As a result, the exact $s$--channel contribution is much larger
% than the NWA calculation, as shown in Fig.~\ref{fig:appronwa}.
% One can see that, for $F=0$ LQ, the deviation from NWA becomes
% important for $M > 250 \GeV$, $\lambda > 0.5$, while for $F=2$ LQ
% this deviation appears already for smaller LQ masses.
%
%..........................................................
%\subsubsection{$s$, $u$ channels and Interference with DIS}
%..........................................................
%
% {\bf{$s$, $u$ channels and Interference with DIS : }}

\begin{figure}[htb]
 \begin{center}
  \hspace*{-2cm}\begin{tabular}{p{0.40\textwidth}p{0.60\textwidth}}
         \vspace*{-6cm}\caption
         { \label{fig:lqsuint}
%              Relative contributions of $s$-channel, $u$-channel and
%              of the interference term (I.T.) with SM DIS, for
%              (a) $S_{0,R}$ ($F=2$) and (b) $S_{1/2,L}$ ($F=0$) LQ. 
%              $\lambda$ has been fixed to 0.5 } &
% ===> EP :
%               Comparison of the contributions of $s$-channel,
%               $u$-channel and interference term (I.T.) with DIS,
%               (convoluted with parton densities and integrated
%               over phase space) for
%               (a) $S_{0,R}$ ($F=2$) and (b) $S_{1/2,L}$ ($F=0$) LQ.
%               $\lambda$ has been fixed to 0.5 } &
% ===> End EP
% ===> RR :
               Contributions of the $s$--and $u$--channel
               processes and of their interference with SM DIS
               for (a) $S_{0,R}$ ($F=2$) and (b) $S_{1/2,L}$ ($F=0$).
               $\lambda$ has been fixed to 0.5, a value
               which is typical for the experimental sensitivity
               in this mass range. } &
 \epsfxsize=0.7\textwidth
 \epsfbox{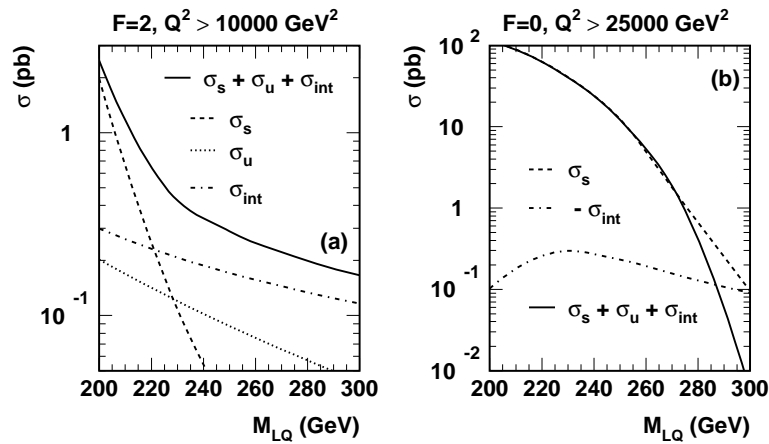}
 \end{tabular}
 \end{center}
 \vspace*{-1.5cm}
\end{figure}

Furthermore, the $u$--channel contribution can also become large 
especially for $F=2$ LQs and $e^+ p$ scattering where 
valence quarks are involved.
A third effect which plays an important role at high $M$ is the
interference with SM $\gamma$ and $Z$ exchange.
Depending on the LQ type, the interference can be constructive
or destructive, and it can get comparable to or greater than
the pure $s$-- and $u$--channel contributions.
In Fig.~\ref{fig:lqsuint} the relative magnitudes of the various
terms in the high $Q^2$ domain are shown for the $S_{0,R}$ ($F=2$) and the
$\tilde{S}_{1/2,L}$ ($F=0$) scalar leptoquarks.
%
% The interference with DIS also plays
% an important role at high $M_{LQ}$. It can be positive or negative
% depending on the LQ type, and can be, in absolute value, comparable
% to or greater than the $s+u$ contributions.
% The contributions of $s$--channel, $u$--channel and
% interference with DIS, after convolution with relevant
% parton density and phase space integration, are shown 
% in Fig.~\ref{fig:lqsuint} for
% $S_{0,R}$ ($F=2$) and $S_{1/2,L}$ ($F=0$) scalar leptoquarks.
%
%------------------------------------------------------
%   FIGURE 6  : Relative contributions of s, u, IT
%------------------------------------------------------
%
% Since the pure DIS cross section is not included, a $Q^2$ cut has been
% applied to evaluate these contributions.
For the $S_{0,R}$, the $u$--channel and the (positive)
interference with SM DIS begin to dominate the $s$--channel 
cross section at
$M > 220 \GeV$. For the $\tilde{S}_{1/2,L}$, the $u$--channel contribution
is negligible, but the (negative) interference becomes important
for $M > 285 \GeV$.

%----------------------------
\subsection{Analysis}
%----------------------------

Resonant leptoquarks decaying into an electron and a quark jet
% ===> TM Dec 16
% lead to signatures which are indistinguishable from the NC DIS
% on an event-by-event basis.
% Statistically, however, the signal could be
lead to signatures which resemble NC DIS events.
However, the signal could be
% ===> End
identified as a resonance peak in the invariant mass distribution,
with events in the peak showing an angular distribution specific
to the spin of the leptoquark state.
For example, the isotropic decay of a scalar LQ in its CM frame leads to a
% ===> TM Dec 16
% flat spectrum in $y$, where $y=1/2 (1 + \cos \theta^*)$,
flat spectrum in $y$, where $y=(1 + \cos \theta^*)/2$,
% ===> End
% End TM
$\theta^*$ being the polar angle of the scattered lepton in this frame.
This is 
% markedly 
remarkably
different from the steeply falling $1 / y^2$ distribution
expected at fixed $x$ for the dominant $t$--channel $\gamma$ exchange
in NC DIS events.
%
%------------------------------------------------------
%   FIGURE 7  : H1 and ZEUS Scatter plots and M spectra
\begin{figure}
 \begin{center}
 \begin{tabular}{cc}
 \epsfxsize=0.4\textwidth
% \epsfbox{h1.ichep98.figscatter.bw.eps}
 \epsfbox{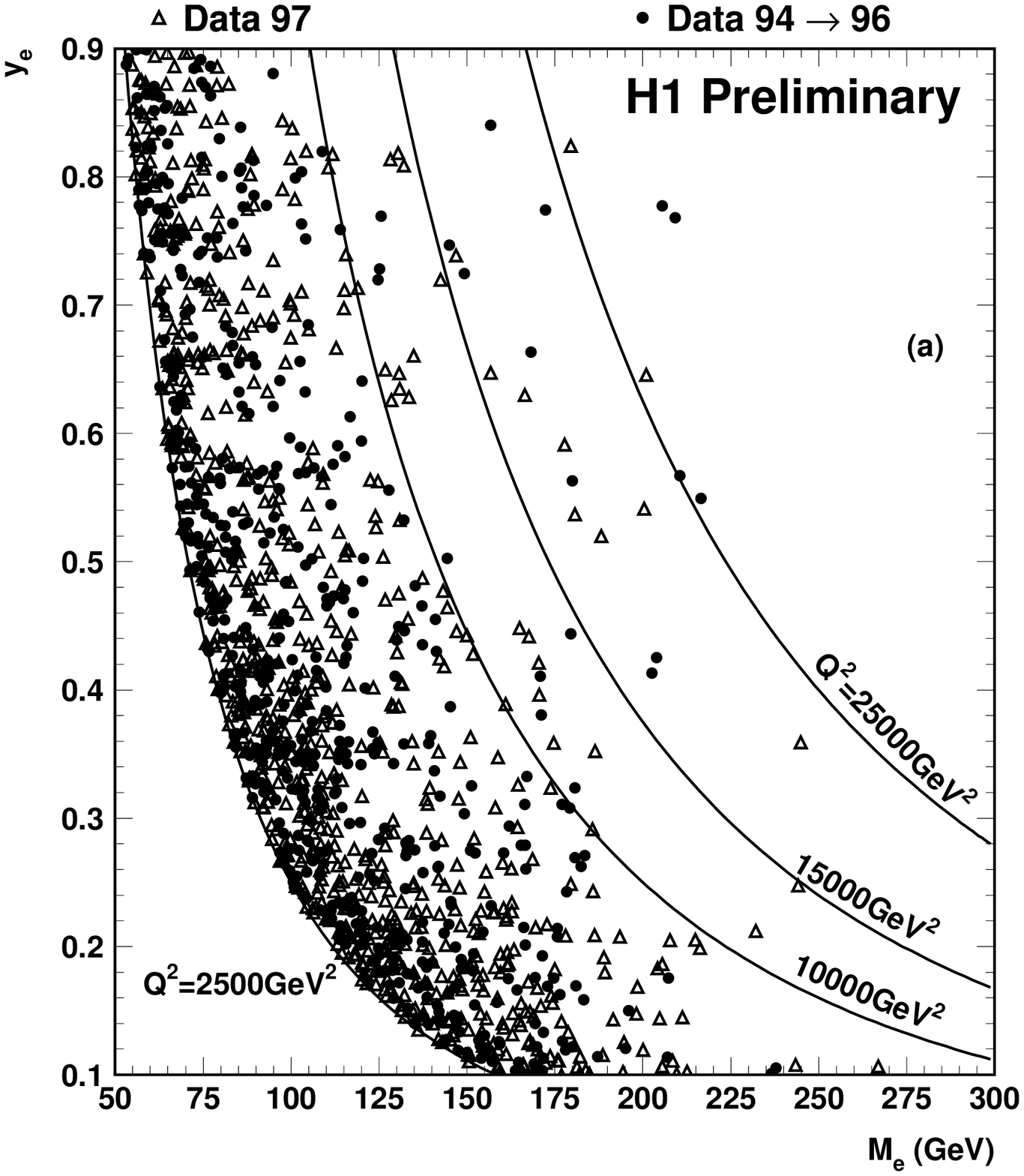}
 &
 \epsfxsize=0.4\textwidth
% \epsfbox{h1.md98.dndm.pap.bw.eps} \\
 \epsfbox{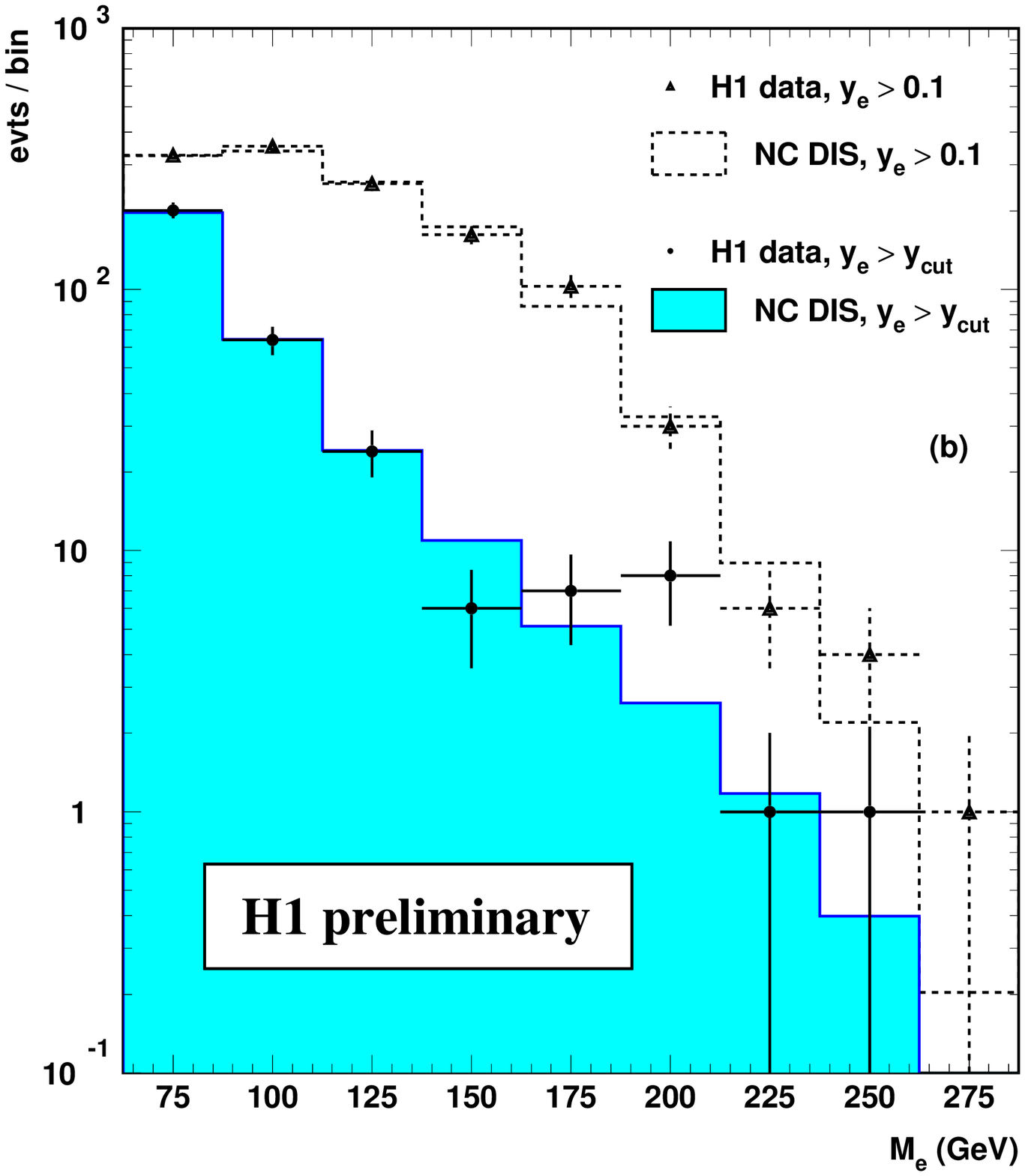} \\
 \epsfxsize=0.4\textwidth
 \epsfbox{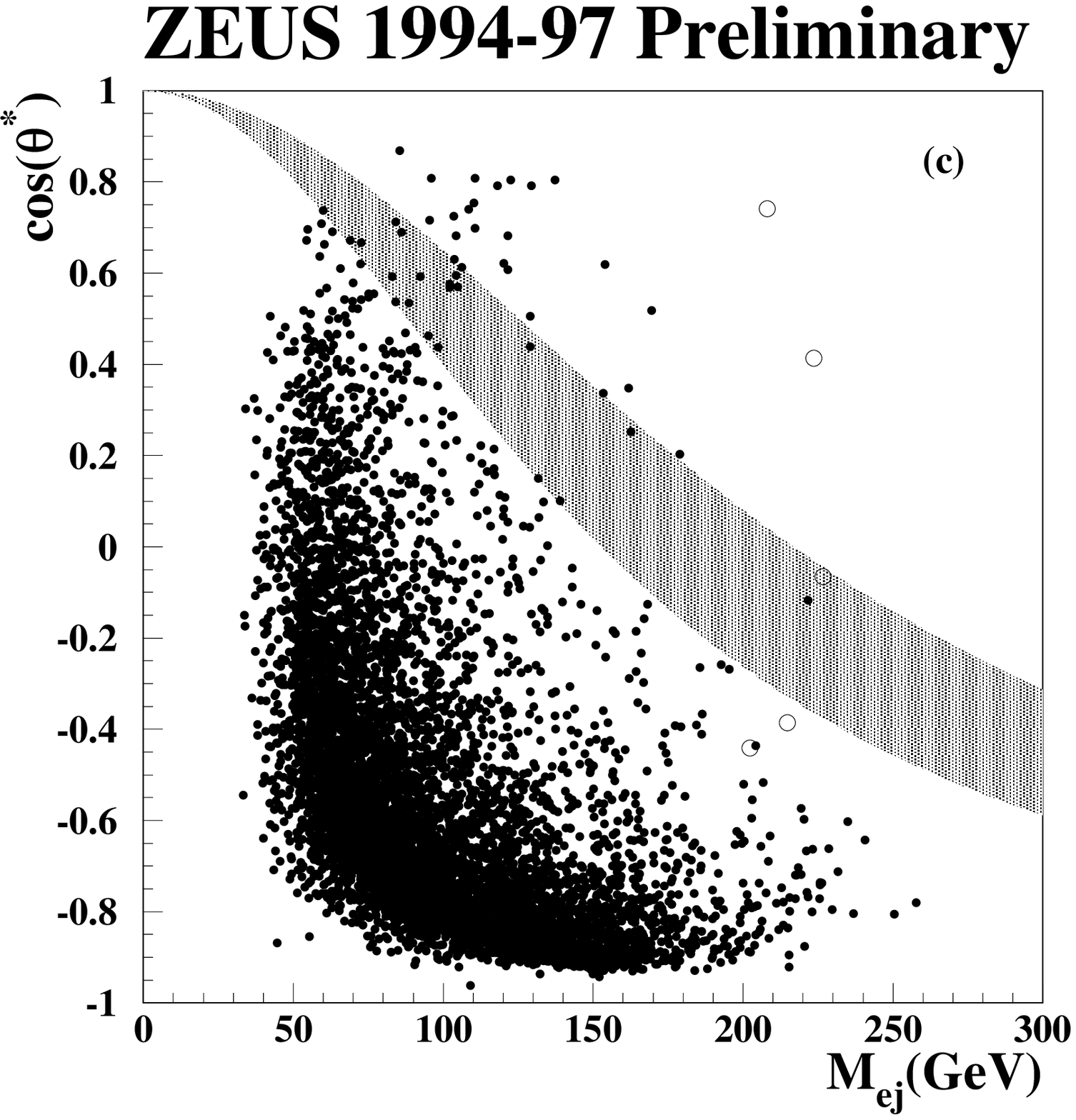}
 &
 \epsfxsize=0.4\textwidth
 \epsfbox{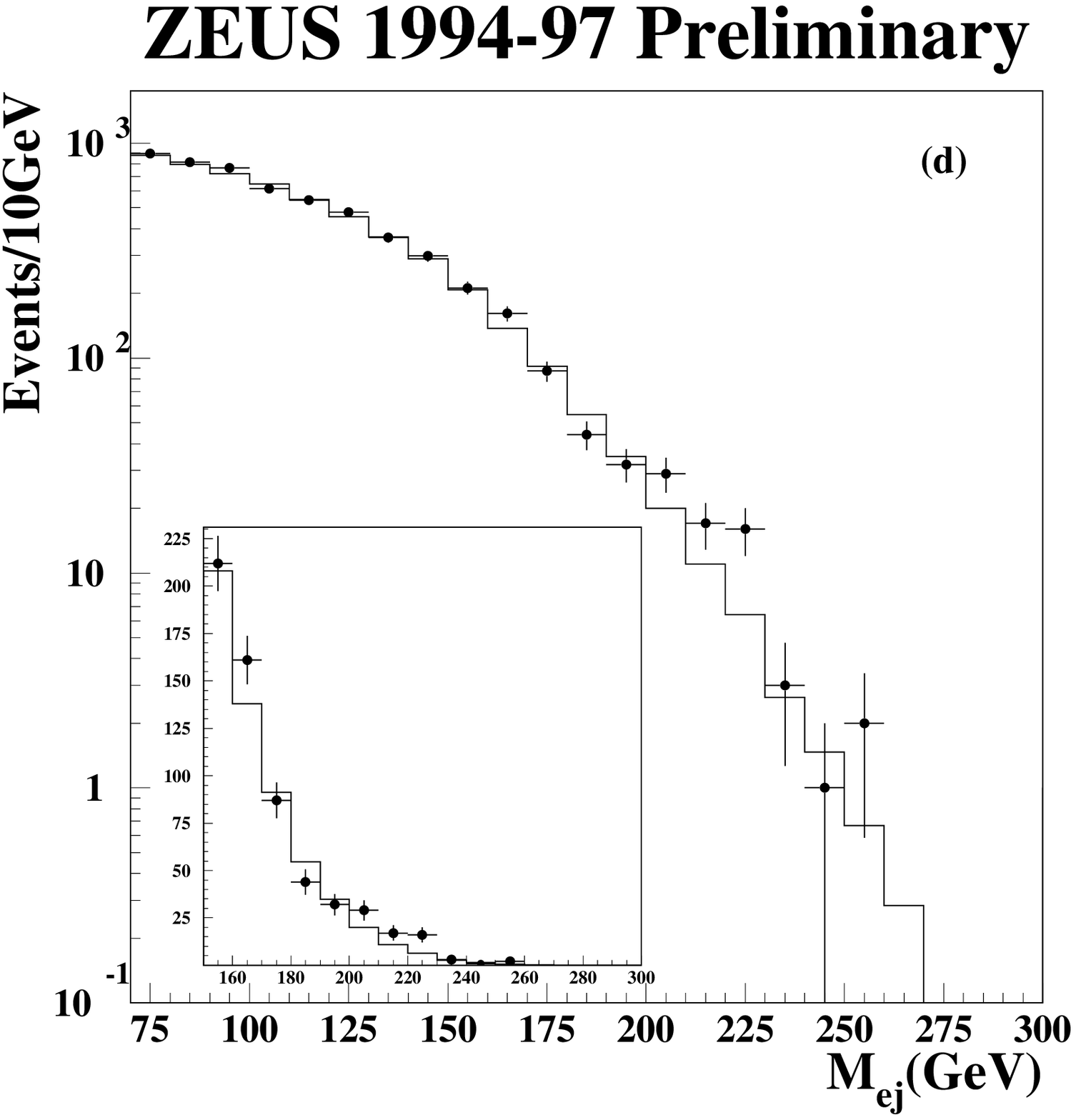}
 \end{tabular}
 \end{center}
 \caption[]{\label{fig:lqplots} 
          High $Q^2$ NC DIS candidates observed by the H1 (a,b) and
          ZEUS (c,d) experiments. 
% ===> TM Dec 16
  (a) Events in the ($M, y$)-plane. Contours of fixed $Q^2$ are shown.
  (b) Mass spectra of the events before (triangles, dotted lines) and
      after (circles, solid lines) the optimal y cut.
  (c) Events in the (cos $\theta^*$, $M_{ej}$)-plane. Shaded is the
      removed region which corresponds to the interface between calorimeters.
      Open circles show the high-$x$ and high-$y$ events reported previously.
  (d) Mass spectra of the events.}
% ===> End
\end{figure}
%------------------------------------------------------
%
Hence, a LQ signal should become more and more prominent at high $y$
(or high $Q^2 = y M^2$).
Consequently, it should be possible to optimize the signal
to NC DIS background ratio by a mass dependent lower cut on $y$
(or equivalently on $\cos \theta^*$).
Note that this optimal cut depends not only on the spin
of the LQ, but also on its fermion number, since the $s$-- and
$u$--channels, which lead to different angular distributions,
contribute differently for $F=0$ and $F=2$.

The NC DIS candidates at $Q^2 > 2500 \GeV^2$ observed by H1 are shown
in Fig.~\ref{fig:lqplots}a in the $(M,y)$--plane.
Here, the full dataset was re-analysed including a new in-situ energy 
calibration for the electrons. This has led to slight migrations
(within originally quoted systematic errors) of individual events
in $(M,y)$.
Fig.~\ref{fig:lqplots}b shows the projected mass spectra,
before and after applying the optimal $y$ cut mentioned
above.
%  which are seen to be well described by the SM expectation.
In the mass window considered in~\cite{H1HIQ2} of total width
$25 \GeV$ around $200 \GeV$, and at $y > 0.4$, 8 events are observed,
while $3.0 \pm 0.5$ are expected from NC DIS. Out of these,
5 events originate from the $1994 - 1996$ data and 3 from the 1997 data.
Hence, the ``mass clustering'' around $\simeq 200 \GeV$ observed previously 
is not confirmed by the 1997 H1 data alone.

% ===> TM Dec 16
ZEUS searched for a resonance in positron-jet mass $M_{ej}$, without
applying any constraints. The $M_{ej}$ does not give the optimal
resolution for LQ, however, it is free from initial state QCD and QED
radiation shifts. The mean reconstructed mass is determined to be within
6\% of the generated value, whereas the peak is on average 4\% lower than
the one obtained from a LQ MC.
% ===> End
Fig.~\ref{fig:lqplots}c shows the distribution of ZEUS NC DIS candidates
in the $( M , \cos \theta^*)$--plane. The fiducial cut applied
on the positron which excludes
the interface region between the central and forward calorimeters
is indicated as the shadowed area.
The projected spectrum of the invariant mass $M_{ej}$ 
is shown in Fig.~\ref{fig:lqplots}d.
It is interesting to note that the outstanding high $Q^2$ events discussed
in~\cite{ZEUSHIQ2} tend to cluster around $M_{ej} \simeq 215 \GeV$.
At $M_{ej} > 200 \GeV$ and $\cos \theta^* > -0.2$, 4 events 
are observed which is in slight excess to the $1.2$ events
expected from NC DIS. 
On the other hand, for the events at $M_{ej} > 200 \GeV$ the
angular distribution of the scattered positron is found to be
similar to that expected from the SM~\cite{ZEUSICHEPLQ}.

%----------------------------
\subsection{Results}
%----------------------------

\begin{figure}
 \begin{center}
 \begin{tabular}{cc}
 \epsfxsize=0.5\textwidth
 \epsfbox{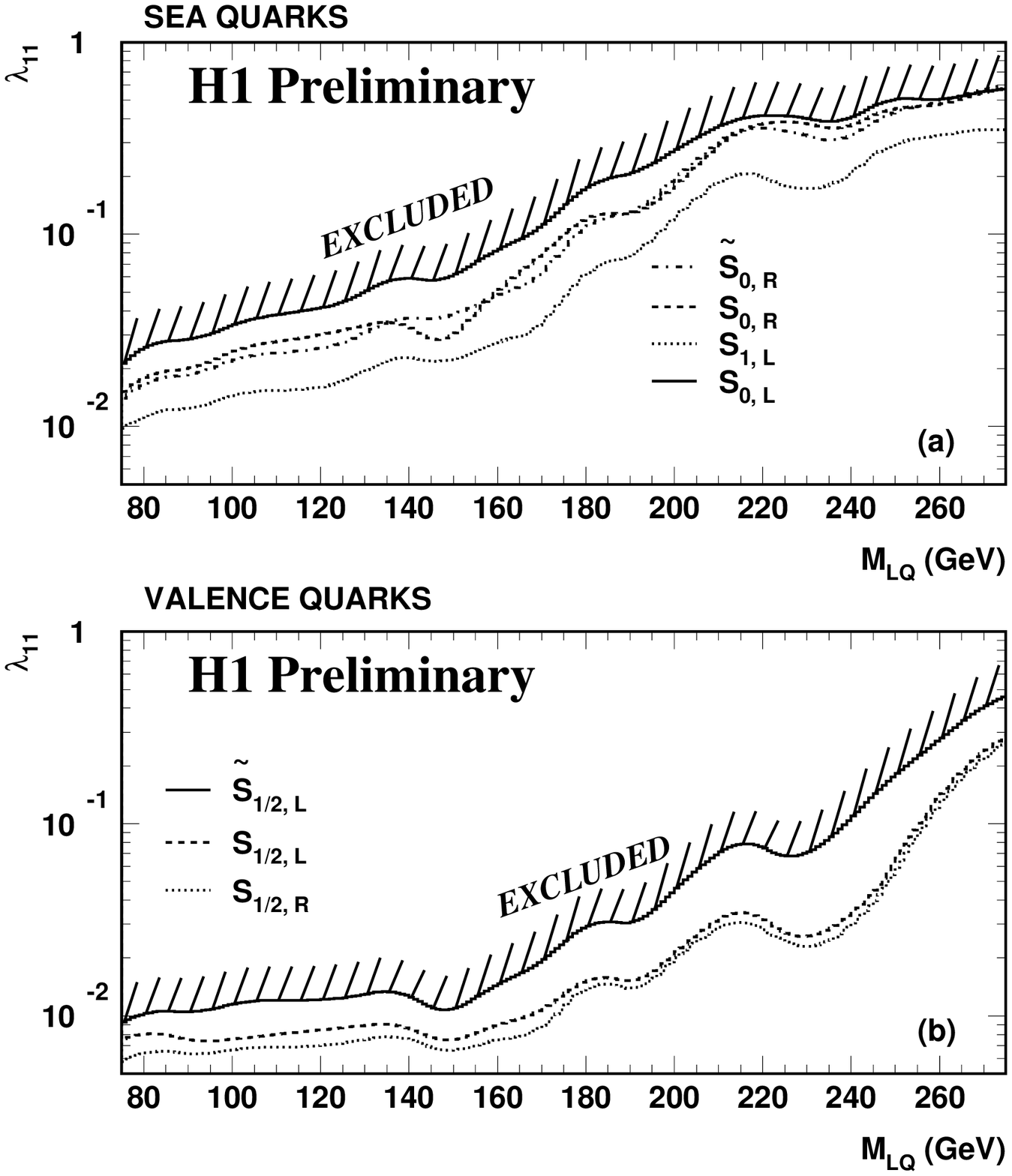}
 &
\epsfxsize=0.5\textwidth
 \epsfbox{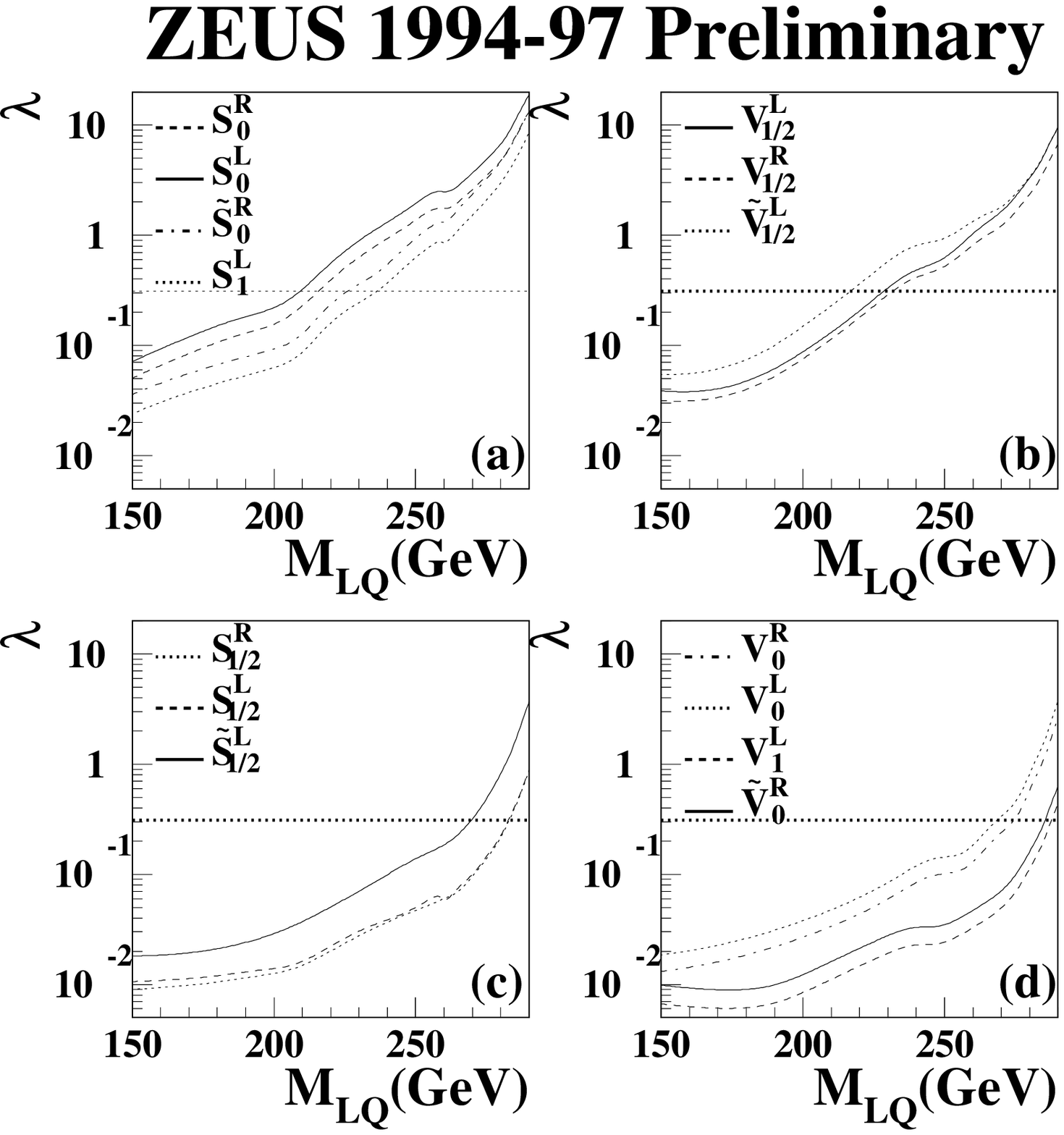}
 \end{tabular}
 \end{center}
 \caption[]{ \label{fig:lqlimits} 
            Left~: H1 exclusion limits at $95 \%$ CL on the Yukawa
            coupling $\lambda$ as a function of the LQ mass (a)
            for $F=2$ and (b) for $F=0$ scalar leptoquarks.
            Right~: analogous limits from ZEUS  (a) for scalars with $F=2$, 
            (b) vectors with $F=2$, (c) scalars with $F=0$,
            and (d) vectors with $F=0$. }
\end{figure}

Assuming that the slight excess of NC events observed at high $Q^2$
is due to a statistical fluctuation, H1 and ZEUS have derived
preliminary limits on the Yukawa coupling $\lambda$ as a function
of the LQ mass, for all LQ species classified by BRW~\cite{BRW}. 
These are shown in Fig.~\ref{fig:lqlimits}.
%
%------------------------------------------------------
%   FIGURE 8  : H1 and ZEUS LQ Limits
%------------------------------------------------------
%
The discrepancy between the H1 and ZEUS bounds for
high mass scalar LQs is due to the use of the NWA approximation
by ZEUS in the calculation of the signal cross section, while
H1 takes into account the LQ width by a Breit-Wigner distribution.
For Yukawa couplings of electromagnetic strength,
$\lambda^2 / 4 \pi = \alpha_{EM}$, the existence of
$F=0$ LQs is excluded up to $M \simeq 275 \GeV$.

Relaxing the BRW assumptions, the H1 Collaboration also obtained
bounds in the plane $\beta$ vs $M$, $\beta$ being the branching
ratio for $LQ \rightarrow e q$, for fixed values of the Yukawa coupling.
The results can be found in~\cite{H1ICHEPLQ} and in these proceedings~\cite{DWATERS}.
For $\beta = 1$, TeVatron experiments exclude first generation LQ masses
below $\simeq 242 \GeV$~\cite{TEVCOMBINED}, independently of
any assumption on $\lambda$. Nevertheless,
for low $\beta$, an important discovery potential
remains for HERA. 
For example, if $\beta = 0.1$, the D$\emptyset$ experiment
excludes masses only up to about $110 \GeV$, while H1 rules out masses
below $255 \GeV$ if $\lambda=0.3$, and $210 \GeV$ 
if $\lambda=0.1$~\cite{H1ICHEPLQ}.
% independently of other possible decay modes
% of the leptoquark~\cite{H1ICHEPLQ}.
%
% ===> EP : Add comment on LEP LQ limits
The DELPHI~\cite{DELPHILQ} and OPAL~\cite{OPALLQ}
experiments at LEP recently performed a search for single LQ
production and exclude LQ masses below $142 \GeV$ for $\lambda = 0.3$, which 
because of the smaller centre of mass energy cannot be
competitive with the sensitivity achieved at HERA.
% ===> End EP

\begin{figure}
 \begin{center}
 \hspace*{-2cm}\begin{tabular}{p{0.50\textwidth}p{0.50\textwidth}}
         \vspace*{-6cm}\caption
         { \label{fig:lqtau}
 Bounds on $\lambda_{3j}$, $j=1,2$, against
 the $S_{1/2,L}$ mass, for several fixed values of $\lambda_{11}$.
 The grey domains are excluded at $95 \%$ confidence level.
 The full curve indicates the most stringent indirect low
 energy limit on $\lambda_{31}$ for $\lambda_{11}=0.3$. } &
 \epsfxsize=0.5\textwidth
 \epsfbox{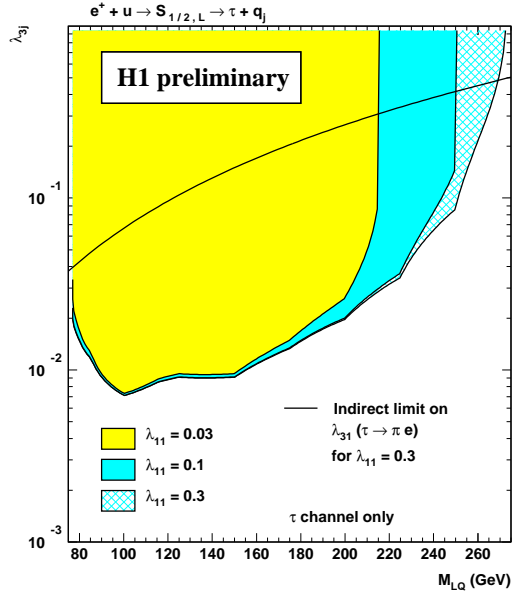}
 \end{tabular}
 \end{center}
 \vspace*{-1cm}
\end{figure}

H1 also carried out a search for Lepton Flavor Violating 
LQs~\cite{H1ICHEPLQ}, looking for LQs decaying into $\mu + q_j$\footnote{
% ===> EP : Add footnote :
Note that the $ep \rightarrow \mu + {\mbox{jet}} + X$ events
observed by H1 and discussed in~\cite{H1MUON}
fail significantly the kinematical constraints that should
hold for the process
$eq \rightarrow LQ \rightarrow \mu q$.}
% ===> End EP 
and $\tau + q_j$ through the coupling $\lambda_{2j}$ 
and $\lambda_{3j}$ respectively.
%
% ===> EP : Add figure for LQ -> tau
% Rejection upper limits on $\lambda_{3j}$ as a function of the
% LQ mass are shown in Fig.~\ref{fig:lqtau} for several fixed
Mass dependent bounds on $\lambda_{3j}$ are shown in
Fig.~\ref{fig:lqtau} for several fixed
values of $\lambda_{11}$. For 
$\lambda_{11}=\lambda_{3j}=0.3$, H1 excludes LQs lighter than
255 GeV, which significantly improves the limit of about $100 \GeV$
obtained from third generation LQ searches carried out at the 
TeVatron~\cite{D03GENE,CDF3GENE}, unless $\lambda$ is 
much smaller.
Indirect limits on $\lambda_{31}$ are also improved
by typically one order of magnitude.
% 
% Especially, the direct and indirect constraints
% on $LQ \rightarrow \tau + jet$ are significantly improved by the
% H1 analysis~\cite{H1ICHEPLQ}.
%------------------------------------------------------
%   FIGURE 9  : Limits for LQ -> tau
%
%------------------------------------------------------

%==========================================================
\section{Searches for R-parity violating squarks}
%==========================================================

\label{sec:rpvsusy}

%----------------------------
\subsection{Phenomenology}
%----------------------------

% ===> RR
The most general superpotential consistent with the gauge symmetry
and supersymmetry of the Minimal Supersymmetric Standard Model (MSSM)
contains Yukawa terms of superfields which violate lepton ($L$)-- and
baryon ($B$)--number. These interactions can give rise to rapid proton
decay. In order to avoid this problem, one introduces additional
discrete symmetries which forbid some or all of the dangerous Yukawa
couplings. A radical solution is achieved by requiring R-parity
conservation where R-parity is defined by $R_p=(-1)^{3B+L+2S}$, $S$
being the spin of a given particle, so that the particle content of the MSSM
can be classified in SM particles with $R_p=+1$ and superpartners
with $R_p=-1$. The $L$-- and $B$--violating Yukawa interactions in the
superpotential being odd under $R_p$ vanish.
However, phenomenologically it is already sufficient to
forbid the $B$--violating interactions and to constrain the 
$L$--violating
terms such that they are consistent with the remaining, much weaker
experimental bounds.

\begin{figure}
 \begin{center}
 \hspace*{-2cm}\begin{tabular}{p{0.50\textwidth}p{0.50\textwidth}}
         \vspace*{-4cm}\caption
         { \label{fig:sqgauge}
%===>RR
 Illustrative example of resonant squark production followed
 by $R_p$-violating and conserving decays into $e^+$ and jets.
%===>end RR
%              Example diagram of resonant squark production
%              followed by ``gauge'' decay of squark into
%              e.g. $e^+$ and jets. 
} &
 \epsfxsize=0.5\textwidth
 \epsfbox{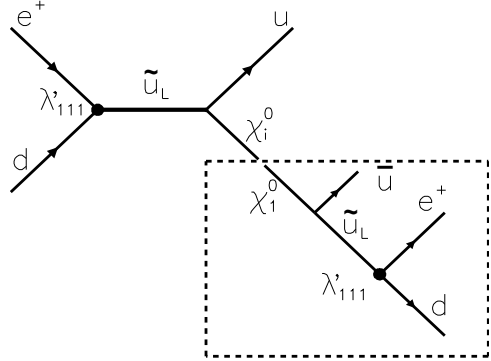}
 \end{tabular}
 \end{center}
 \vspace*{-1cm}
\end{figure}

In such a scenario one can have the term
$\lambda'_{ijk} L_i Q_j \bar{D}_k$
in the superpotential, where $L_i$ and $Q_j$ are the lepton and quark
$SU(2)_L$-doublet superfields, respectively, $\bar{D}_k$ is the
down-quark $SU(2)_L$-singlet superfield, and $\lambda'_{ijk}$
denote the dimensionless Yukawa couplings. The indices $i,j,k$ refer to
the fermion generations.
Most interestingly at HERA, 
the $R_p$--violating couplings with
$i=1$ can give rise to resonant squark production~\cite{RPVIOLATION}
\footnote{Further studies of $R\hspace{-0.2cm}/_p$-processes at HERA
can be found in~\cite{H1RPV96,DRPEREZ,HERAWRK96}.}.
In $e^+p$ scattering, the possible production channels are
  $e^+ \bar{u}_j \rightarrow \bar{\tilde{d}}_{Rk}$ and
  $e^+ d_k \rightarrow \tilde{u}_{Lj}$.
In particular valence $d$ quarks allow to probe 
$\tilde{u}_{Lj}$ squarks and $\lambda'_{1j1}$ couplings.
The present indirect bounds on $\lambda'_{1j1}$ and 
$M_{\tilde{q}}$~\cite{DRBOUNDS}
leave an interesting search window for HERA with the exception
of $\lambda'_ {111}$ in which case the limit from neutrinoless
$\beta \beta$ decay
precludes the observation of effects at HERA.
The possible squark interpretations of the observed
excess of events in NC DIS reported in~\cite{H1HIQ2,ZEUSHIQ2}
have been studied in detail 
in~\cite{ALTARELLI,CHOUDHURY,DREINER97,KALINOW}.
In the following, we consider scenarios with a single nonvanishing
coupling $\lambda'_{1j1}$.
Furthermore, we assume that gluinos are
heavier than squarks, so that $\tilde{q} \rightarrow \tilde{g}+q$
is kinematically forbidden. 
% ===> EP :
% The resonantly produced squark then decays
% most probably 
The resonantly produced squark can then decay
% ===> End EP
into $e+q$ via the
$R\hspace{-0.25cm}/_p$-coupling $\lambda'_{1j1}$ resembling a scalar
leptoquark, and into a neutralino $\chi_i^0$ or a
chargino $\chi_i^\pm$ plus a quark via $R_p$-conserving gauge
interactions. In the latter cases,
the final state is determined by the subsequent neutralino and chargino
decays. 
%
% ===> EP :
%    I change because this seems to say that
%    heavier chis ONLY undergo ``gauge'' decays, while
%    \Rp decays are also possible. 
% While the heavier states $\chi_{i>1}$ are expected to
% cascade down into the lightest states, these finally decay
% into SM-particles, e.g.
% $\chi^0_1 \rightarrow e^\pm q \bar{q}'$ or $\nu q \bar{q}$,
% involving again the coupling $\lambda'_ {1jk}$. Similar
% decay channels exist for the $\chi^+_1$ with the exception of the
% $e^-$-mode.
%
The heavier gauginos may cascade down to the lightest one,
while the lightest neutralino may finally
decay into
$e^{\pm} q \bar{q}'$ or $\nu q \bar{q}$,
involving again the coupling $\lambda'_ {1j1}$. Similar
\Rp\-- decay channels exist for the $\chi^+$ with the exception of the
$e^-$-mode. 
% ===> End EP
%
An illustrative example is shown in Fig.~\ref{fig:sqgauge}.
In~\cite{H1RPV96}, a classification is given of all possible final
states in terms of several distinguishable event topologies.
Recently, the SUSYGEN~\cite{SUSYGEN} package
has been extended to include $R\hspace{-0.25cm}/_p$-SUSY
processes at HERA. The new version which will be released shortly
allows a complete event generation for all possible
channels.
% ===> End RR
%
%------------------------------------------------------
%   FIGURE 10  : Squark production and decay
%
%------------------------------------------------------
%
% All possible final states have been classified into several
% distinguishable event topologies in~\cite{H1RPV96}.
% Recently, the SUSYGEN~\cite{SUSYGEN} package has been extended
% to study \Rp SUSY processes at HERA, thus allowing a complete 
% event generation of all kinds of decays. This extension should be part
% of the next release of SUSYGEN.

%----------------------------
\subsection{Analysis}
%----------------------------

\begin{figure}
 \begin{center}
 \hspace*{-2cm}\begin{tabular}{p{0.40\textwidth}p{0.60\textwidth}}
         \vspace*{-4cm}\caption
         { \label{fig:dndms3}
%===>RR
  Invariant mass spectrum of $e^\pm$ + multijet final states
  for data (symbols) and NC DIS expectation (histogram).
%===>end RR
%              Mass spectrum for $e$ + multijet final states
%             for data (symbols) and NC DIS expectation (histogram).
 } &
 \epsfxsize=0.6\textwidth
 \epsfbox{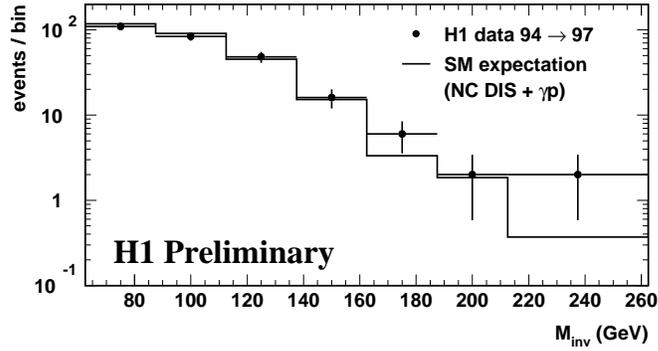}
 \end{tabular}
 \end{center}
 \vspace*{-1cm}
\end{figure}

% We concentrate here on final states containing $e^{\pm}$ and several
% jets, arising e.g. from the pattern $\tilde{q} \rightarrow \chi^0_1 + q$
% followed by the \Rp decay of the $\chi^0_1$.
% This topology has been analysed by H1 Collaboration~\cite{H1ICHEPSQ}.
% The main selection criteria require an $e^{\pm}$ candidate reconstructed
% at high $y_e$ (because the $e$ appears late in the cascade decay chain, hence
% carrying only a part of the $\chi$ energy), together with two quite forward
% jets. The invariant mass spectrum of the 289 selected events is compared
% in Fig.~\ref{fig:dndms3} with the prediction from SM, from which
% $285.7 \pm 28.0$ events are expected, and a very good agreement
% is observed.
%
% ===> RR : Modified text :
The H1 Collaboration \cite{H1ICHEPSQ} analysed final states containing
an $e^\pm$ and several jets.
Since the lepton appears late in the cascade decay chain, it is expected
to be degraded in energy. Consequently,
the main selection criteria require an $e^\pm$ candidate reconstructed
at high $y_e$ together with two forward jets.
The invariant mass spectrum of the 289 events selected
is shown in Fig.~\ref{fig:dndms3} and compared 
with the distribution of the $285.7 \pm 28.0$
events expected in the SM. The agreement is good.
In the ``wrong sign''
channel, $e^+ + p \rightarrow e^- +$ multijet, one
candidate is observed, while $0.49 \pm 0.2$ are expected.
% ===> End RR
%
%------------------------------------------------------
%   FIGURE 11  : H1 Mass spectrum for e + jets final states
%------------------------------------------------------
%
% Note that in the ``wrong sign'' channel 
% $e^+ + p \rightarrow e^- + {\mbox{multijet}} + X$, one candidate
% is observed, well compatible with the SM expectation of $0.49 \pm 0.2$.

%----------------------------
\subsection{Results}
%----------------------------

\begin{figure}
 \begin{center}
 \hspace*{-2cm}\begin{tabular}{p{0.40\textwidth}p{0.60\textwidth}}
         \vspace*{-7cm}\caption
         { \label{fig:sqlimits}
             $95 \%$ CL upper limits on the coupling $\lambda'_{1j1}$
             as a function of the squark mass, for different masses and
%===>RR
composition of the lightest neutralino $\chi^0_1$.
%===>end RR
%             mixtures of the $\chi^0_1.
             The most stringent indirect limits from neutrinoless
             $\beta \beta$ decay and atomic parity violation
             are also indicated. } &
 \epsfxsize=0.5\textwidth
 \epsfbox{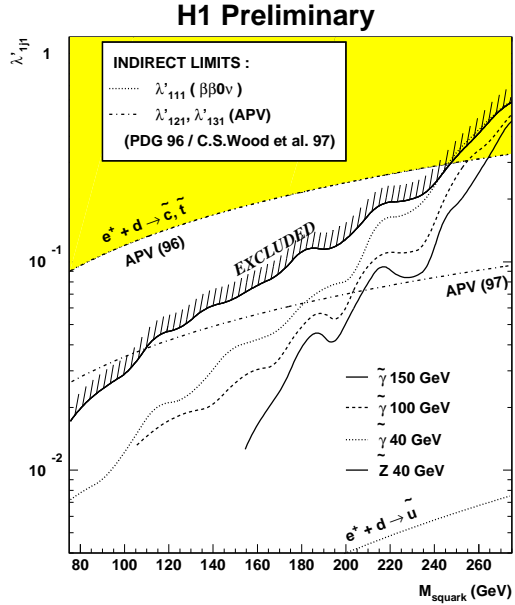}
 \end{tabular}
 \end{center}
 \vspace*{-1cm}
\end{figure}

% The ``leptoquark-like'' channel is combined with the 
% $e + {\mbox{multijet}}$ channel
% to set mass dependent limits on the coupling $\lambda'_{1j1}$ at
% $95 \%$ CL.
% Four set of MSSM parameters $\mu$, $M_2$ and $\tan \beta$ have been chosen,
% leading to a $40 \GeV$ $\chi^0_1$ dominated by its $\tilde{\gamma}$
% or by its $\tilde{Z}$ component, to a $100 \GeV$ $\tilde{\gamma}$-like
% $\chi^0_1$ and to a $150 \GeV$ $\tilde{\gamma}$-like $\chi^0_1$.
% The values of these parameters fix the $R_p$ conserving widths of the relevant
% SUSY particles, and thus the relative contributions of the different 
% channels are known for a given $\lambda'_{1j1}$ value.
%
% ===> RR : modified text :
Combining the LQ-like channel $e^+ +$ jet with the
$e^\pm +$ multijet channels, 
one can put mass-dependent limits on the couplings $\lambda'_{1j1}$.
To this end, four different sets of values have been chosen
for the MSSM parameters $\mu$, $M_2$ and $\tan \beta$, leading to
four typical scenarios for the lightest neutralino~: photino-like states
with mass 40, 100, and 150 GeV, and a zino-like state with mass 40 GeV.
Given $\lambda'_{1j1}$ and the squark mass, the relevant
widths and branching ratios are fixed. 
%
% ===> EP : Refer to the figure !
% This H1 analysis
% As can be seen in Fig.~\ref{fig:sqlimits},
The H1 analysis shown in Fig.~\ref{fig:sqlimits}
% ===> End EP
excludes squark masses up to 262 GeV for $\lambda'_{1j1} \simeq 0.3$.
As expected,
in the case of $\lambda'_{111}$, the H1 bound is not competitive
with the best indirect constraint coming from the non-observation of
neutrinoless $\beta \beta$ decay. However, for $\lambda'_{121}$ and
$\lambda'_{131}$, the best indirect limits from atomic parity
violation
are improved by the H1 direct bounds in a large region of the parameter
space.
Especially, for large $\chi_1^0$ masses, the upper bounds on
$\lambda'_{121}$ and $\lambda'_{131}$ are decreased by as much as a
factor 4.

%
%------------------------------------------------------
%   FIGURE 12  : H1 Limits on Rp-violating SUSY
%------------------------------------------------------
%
% For electromagnetic coupling strengths, squark masses up to $262 \GeV$
% are excluded by this analysis.
% For $\lambda'_{111}$ coupling, H1 limits are not competitive with the best
% indirect limits coming from the non observation of neutrinoless 
% double $\beta$ decay. However, for $\lambda'_{121}$ and $\lambda'_{131}$,
% the most stringent indirect limits come from atomic parity violation
% and are overseeded by H1 direct limits in a large range of parameter 
% space. 
% Especially, for large $\chi^0_1$ masses, HERA limits extend the
% excluded domain
% on $\lambda'_{121}$, $\lambda'_{131}$ by a factor $\simeq 4$.  
%
% First generation LQ searches at TeVatron constrain \Rp squarks,
% but these limits are easily evaded as soon as 
% $\beta (\tilde{q} \rightarrow e + q)$ is small, which arises
% naturally in \Rp SUSY. Recently, TeVatron experiments have performed
%a specific \Rp SUSY analysis~\cite{D0RPV,CDFRPV} looking at like-sign electrons
% plus jets originating from the decay of the pair produced squarks,
% and excluded squark masses below $252 \GeV$, assuming five
% degenerate squarks. If one squark is much lighter than the others
% (e.g. one state of the stop), limits fall around $130 - 150 \GeV$.
%
% ===> RR : modified text :
The $R\hspace{-0.25cm}/_p$ squark scenarios considered above are
also constrained by the first generation LQ searches at the TeVatron.
However, these bounds become
rather weak when $\beta (\tilde{q} \rightarrow e + q)$ is small, 
a possibility which exists
% ===> EP : 
% as may be the case
% as is generally the case
% ===> End EP
due to the presence of $R_p$-conserving decays. Recently, the
TeVatron experiments have performed a dedicated \Rp\
search \cite{D0RPV,CDFRPV} analysing final states with
like-sign electrons plus jets. These may
originate from the decay of pair-produced squarks. Assuming five
degenerate
squark flavours, squark masses below 252 GeV are excluded.
% This bound decreases to 130 -- 150 GeV if
% one squark (e.g.the lightest stop) is much lighter than 
% the others.
If one squark (e.g. the lightest stop) is much lighter than
the others, the cross section is significantly smaller.
The corresponding mass bound is estimated~\cite{H1ICHEPSQ}
to decrease to about $150 \GeV$.

%=========================================================
\section{Conclusions}
%=========================================================

Results on neutral and charged current
Deep Inelastic Scattering at high momentum transfer 
derived from the $e^+p$
data taken by the H1 and ZEUS experiments in 1994--1997,
have been presented and compared to the Standard Model
expectations.
The differential cross sections in terms of $Q^2$ are found to
be in good agreement with the SM predictions. A slight
excess at high $Q^2$ observed in the 1994--1996 data
still persists with the 1997 data included, but with less
significance.

Using the same data, a search for leptoquarks and for squarks 
with R-parity violating couplings has been performed. No evidence 
was found for the existence of such species. The resulting bounds 
on leptoquark/squark masses and Yukawa couplings to lepton-quark pairs 
improve the constraints from other colliders and from low energy
experiments in important regions of the parameter space.

%%%
\ack
TM would like to thank the British Council, Collaborative Research Project
TOK/880/11/15. RR acknowledges support by the Bundesministerium
f\"ur Bildung, Wissenschaft, Forschung und Technologie, Bonn, Germany,
Contract 05 7WZ91P (0).

%%%

%=========================================================
\section{References}
%=========================================================

\end{document}